\documentstyle[11pt,newpasp,twoside,epsf]{article}
\begin{document}
%
%
\title{{\tt SAURON} Observations of Disks in Early-Type Galaxies}
\author{M.\ Bureau, Y.\ Copin, E.\ K.\ Verolme, P.\ T.\ de Zeeuw}
\affil{Sterrewacht Leiden, Postbus 9513, 2300~RA Leiden, Netherlands}
\author{R.\ Bacon, Eric Emsellem}
\affil{CRAL-Observatoire, 9 Avenue Charles-Andr\'{e}, 69230
  Saint-Genis-Laval, France}
\author{Roger L.\ Davies, Harald Kuntschner}
\affil{Physics Department, University of Durham, South Road, Durham
   DH1~3LE, United Kingdom}
\author{C.\ Marcella Carollo}
\affil{Department of Astronomy, Columbia University, 538 West 120th Street, 
   New York, NY~10027, U.\ S.\ A.}
\author{Bryan W.\ Miller}
\affil{Gemini Observatory, Casilla 603, La Serena, Chile}
\author{G.\ Monnet}
\affil{European Southern Observatory, Karl-Schwarzschild Strasse 2, 
   D-85748 Garching, Germany}
\author{Reynier F.\ Peletier}
\affil{Department of Physics and Astronomy, University of
   Nottingham, University Park, Nottingham NG7~2RD, United Kingdom}
%
%
\begin{abstract}
  We briefly describe the {\tt SAURON} project, aimed at determining the
  intrinsic shape and internal dynamics of spheroids. We focus here on the
  ability of {\tt SAURON} to identify gaseous and stellar disks and to measure
  their morphology and kinematics. We illustrate some of our results with
  complete maps of NGC~3377, NGC~3623, and NGC~4365.
\end{abstract}
%
%
\section{The {\tt SAURON} Project}
We are carrying out a study to determine the distribution of intrinsic shapes
of early-type galaxies and bulges and to constrain the range of internal
velocity, age, and metallicity distributions realized in luminous galaxies. Of
interest are the observed dichotomy of central cusp slopes, the dynamical role
of central massive black holes, and the relation between the kinematics and
stellar populations.

\noindent Such goals require the use of wide-field spectroscopy with complete spatial
coverage, a large bandpass, and sufficient spectral resolution. {\tt SAURON} was
designed specifically to meet these goals. In its low spatial resolution
mode, it has a $41\arcsec\times33\arcsec$ field-of-view sampled at 0\farcs94
and a $4800-5400$~\AA\ wavelength range with $3.6$~\AA\ resolution (90~km~s$^{-1}$
instrumental dispersion). The instrument is fully described in Bacon et al.\ 
(2001).

\noindent For the {\tt SAURON} project, we are observing a {\em representative} sample
of spheroids covering the full range of environment, nuclear cusp slope,
rotational support, and apparent flattening. The sample was chosen from a
complete list of galaxies to populate homogeneously six ($\epsilon$, $M_B$)
planes. These planes, representing E/S0/Sa and field/cluster galaxies, are
illustrated in Figure~1. The data are reduced in a uniform manner using a
pipeline specifically developed for {\tt SAURON}.

\begin{figure}
\plotone{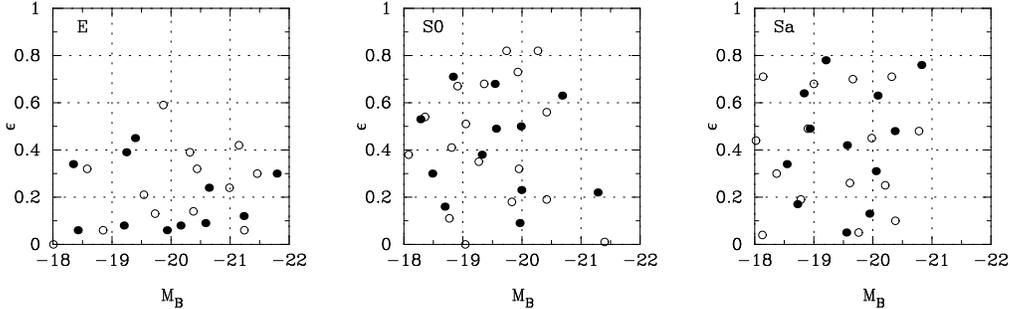}
\caption{The {\tt SAURON} representative sample of E, S0, and Sa
  galaxies. {\em Filled circles:} Cluster galaxies.  {\em Empty circles:}
  Field galaxies.}
\end{figure}
%
%
\section{{\tt SAURON} Observations of Disks}
The wavelength range covered by {\tt SAURON} for nearby galaxies includes many
important emission and absorption lines, such as H$\beta$, [OIII], Mgb, Fe1,
and [NI]. The emission lines allow us to meaure the total intensity and mean
velocity of the ionized gas, but our main goal is to measure the full stellar
line-of-sight velocity distribution from the absorption lines (i.e.\ $v$,
$\sigma$, $h_3$, $h_4$) and to measure line-strength indices. Below, we
present examples of a gaseous disk (NGC~3377) and a stellar disk (NGC~3623)
embedded in a larger spheroid. We also show line-strength maps for a galaxy
harboring a central kinematically decoupled component (NGC~4365). Disks,
asymmetries, and evidence of triaxiality are all easily detected (see de Zeeuw
et al.\ 2001).
%
%
\subsection{A Barred Gaseous Disk in NGC~3377}
NGC~3377 is a large E6 galaxy in the Leo group, previously thought to be
axisymmetric. Figure~2 shows the [OIII] total intensity and velocity field
obtained with {\tt SAURON}. The ionized gas distribution is strongly
non-axisymmetric, resembling a bar. The gaseous kinematics also suggests a
bar-like perturbation, showing minor-axis rotation and a twist of the
kinematic major axis. Evidence of triaxiality is also seen in the {\tt SAURON}
stellar velocity field of NGC~3377, which shows significant minor-axis
rotation (Bacon et al.\ 2001).

\begin{figure}
\plotone{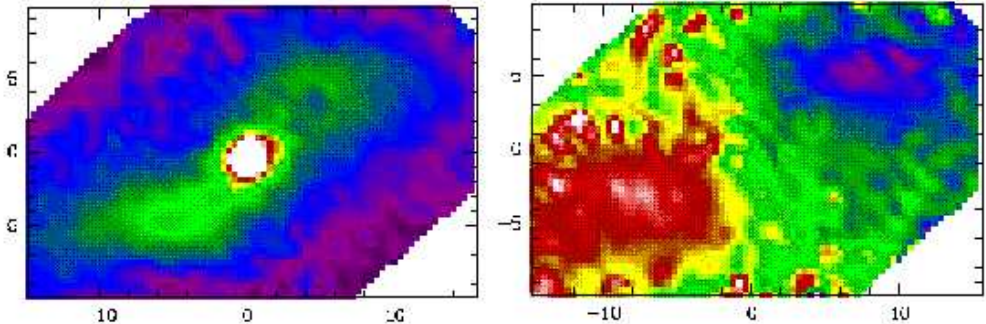}
\caption{The {\tt SAURON} [OIII] total emission line intensity ({\em left})
  and velocity field ({\em right}) for the E6 galaxy NGC~3377. The scales are
  in arcseconds and velocities extend from -140 to 140~km~s$^{-1}$.}
\end{figure}
%
%
\subsection{A Central Stellar Disk in NGC~3623}
NGC~3623 (M~65) is another highly-inclined galaxy in the Leo group, but it is
of much later type than NGC~3377, SABa(rs). It is part of the Leo triplet with
NGC~3627 and NGC~3628 but does not appear to be interacting. NGC~3623's
kinematics has barely been studied and the {\tt SAURON} observations provide the
first glimpse of its dynamics. Figure~3 shows the reconstructed luminosity
distribution, velocity field, and velocity dispersion field obtained with
{\tt SAURON}. They encompass most of the bulge. The large-scale velocity field
reveals minor-axis rotation, in agreement with the presence of a bar. In
addition, a quasi edge-on disk is present in the center, where the isovelocity
contours flatten out abruptly. This disk appears as a central depression in
the velocity dispersion field, and its effects are also seen in the $h_3$ and
$h_4$ maps (not shown).

\begin{figure}
\plotone{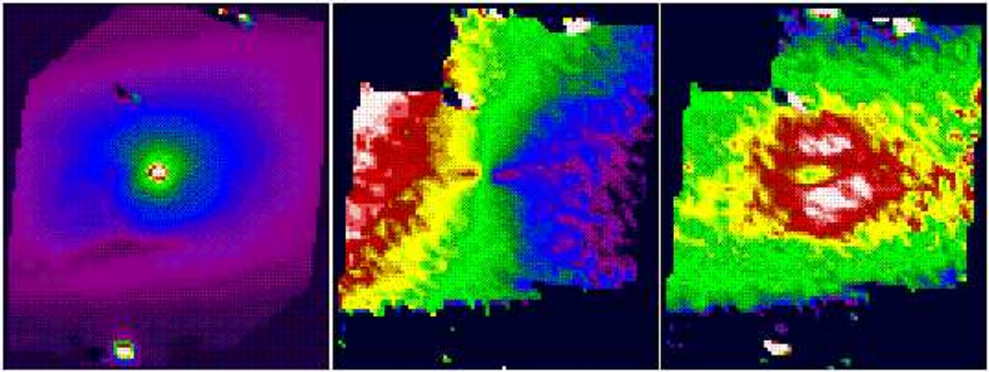}
\caption{The {\tt SAURON} reconstructed luminosity distribution ({\em left}),
  velocity field ({\em center}), and velocity dispersion field ({\em right})
  for the SABa galaxy NGC~3623. Each field is approximately
  $48\arcsec\times62\arcsec$ and represents the mosaic of two partially
  overlapping {\tt SAURON} pointings. Velocities range from -180 to
  180~km~s$^{-1}$ and dispersions from 80 to 180~km~s$^{-1}$.}
\end{figure}
%
%
\subsection{The Kinematically Decoupled Core of NGC~4365}
NGC~4365 is a large E3 galaxy in Virgo. It was known to possess a
kinematically decoupled core, rotating perpendicularly to the outer parts (in
projection; see Surma \& Bender 1995). Figure~4 presents maps of the
reconstructed luminosity distribution, velocity, and velocity dispersion, as
well as the Mgb and H$\beta$ line-strength maps in the Lick/IDS system. The
{\tt SAURON} data reveal the morphology, kinematics, and line-strengths of the stars
over most of the body of the galaxy. The two components with misaligned
angular momenta are easily recovered and suggest that the galaxy is triaxial
or at least that it departs significantly from axisymmetry. The velocity
dispersion and line-strengths follow the light distribution and show no sign
of the central component. The line-strengths suggest a constant
luminosity-weighted age of $\approx14$~Gyr across the galaxy, accompanied by a
decrease of the metallicity with radius; only the very center may be 2--3~Gyr
younger. These results imply that the kinematic configuration is stable over a
Hubble time and that the entire galaxy experienced a similar star formation
history, most likely involving considerable gaseous dissipation (see Davies et
al.\ 2001 for more details).

\begin{figure}
\plotone{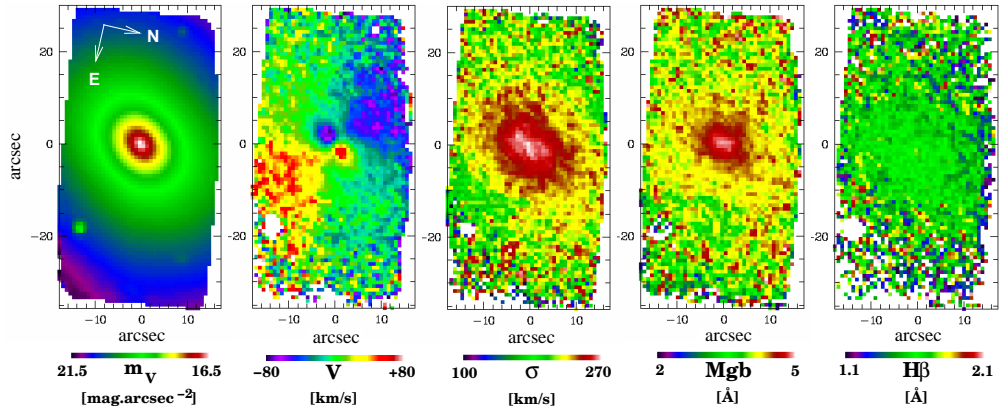}
\caption{The {\tt SAURON} reconstructed luminosity distribution, velocity,
  velocity dispersion, Mgb, and H$\beta$ maps for the E3 galaxy NGC~4365. The
  data represent the mosaic of two partially overlapping {\tt SAURON}
  pointings.}
\end{figure}
%
%

%
\end{document}